\documentclass[prb,aps,nobalancelastpage,amssymb,twocolumn,citeautoscript]{revtex4}

\usepackage{graphicx}

\begin{document}

\title{Dimensionality-driven spin-flop transition in quasi-one-dimensional PrBa$_2$Cu$_4$O$_8$}

\author{ Xiaofeng Xu$^{1,2}$, A. Carrington$^1$, A. I. Coldea$^1$, A. Enayati-Rad$^1$, A. Narduzzo$^3$, S. Horii$^4$ and N. E. Hussey$^1$}

\affiliation{$^1$H. H. Wills Physics Laboratory, University of Bristol, Tyndall Avenue, BS8 1TL, United Kingdom}
\affiliation{$^2$Department of Physics, Hangzhou Normal University, Hangzhou 310036, China}
\affiliation{$^3$Department of Physics, University of Bath, Bath, BA2 7AY, United Kingdom}
\affiliation{$^4$Department of Environmental Systems Engineering, Kochi University of Technology, Tosayamada, Kochi 782-8502, Japan}

\date{\today}

\begin{abstract}
In the quasi-one-dimensional cuprate PrBa$_2$Cu$_4$O$_8$, the Pr cations order antiferromagnetically at 17 K in zero field. Through a combination of magnetic susceptibility, torque magnetometry, specific heat and interchain transport measurements, the anisotropic temperature-magnetic field phase diagram associated with this ordering has been mapped out. A low-temperature spin-flop transition in the Pr sub-lattice is found to occur at the same magnetic field strength and orientation as a dimensional crossover in the ground state of the metallic CuO chains. This coincidence suggests that the spin reorientation is driven by a change in the anisotropic Rudermann-Kittel-Kasuya-Yosida (RKKY) interaction induced by a corresponding change in effective dimensionality of the conduction electrons.
\end{abstract}

\maketitle

\section{Introduction}
\label{intro}

The complex interplay between charge and spin degrees of freedom plays a key role in the physics of correlated electron systems. In the majority of cases, this interplay is manifest in a response of the charge sector to changes in the (localized) spin background. Notable examples include lightly doped, insulating cuprates, where a metamagnetic \cite{Thio88} or spin-flop transition \cite{Lavrov03} of the Cu moments is accompanied by a marked change in the hopping conductivity of the doped holes, the organic salt $\lambda$-(BETS)$_2$FeCl$_4$ where an insulator-metal transition coincides with a field-induced ordering of the Fe$^{3+}$ spins \cite{Brossard98} and the frustrated metallic antiferromagnet 2{\it H}-AgNiO$_2$, where a cascade of magnetic transitions on the localized Ni sites induces corresponding changes in the mobility and Fermiology of the itinerant $d$-electrons.\cite{Coldea09}

Here we demonstrate a rare, if not unique, counter example of a transformation of the spin configuration on a magnetic sub-lattice driven by a fundamental change in the ground state of the mobile carriers. The system in question is PrBa$_2$Cu$_4$O$_8$ (Pr124), a non-superconducting analog of the underdoped superconducting cuprate YBa$_2$Cu$_4$O$_8$ (Y124), with edge-sharing, highly conducting CuO double chains oriented along the $b$-axis, sandwiched between $ab$-plane CuO$_2$ bilayers (see inset in Fig.~\ref{Fig1}).\cite{Terasaki96, McBrien02, Hussey02} Due to strong hybridization of the 2$p$-orbitals of oxygen with the Pr 4$f$-orbitals,\cite{FehrenbacherRice} the CuO$_2$ plane carriers localize and order antiferromagnetically around $T_{\rm N}$(Cu) $\sim$ 220 K.\cite{Li99} The Pr cations also order antiferromagnetically at $T_{\rm N}$(Pr) = 17 K, \cite{Yang97} a temperature that is one order of magnitude higher than the $T_{\rm N}$ values found in all the other rare-earth element containing layered cuprates.\cite{Ku01}

Early neutron diffraction measurements on Pr124 suggested that the Pr ions order collinearly along the $c$-axis. \cite{Li99} A similar arrangement of spins was originally proposed in the single-chain compound PrBa$_2$Cu$_3$O$_7$ (Pr123) \cite{Li89} but later it was suggested that the spins are actually tilted out of the plane. \cite{Longmore96, Boothroyd97} These latter measurements however could not resolve the direction ($a$- or $b$-axis or both) of the tilting. One initial motivation for our study was to shed light on the nature of the Pr ordering in Pr124 by carrying out a detailed study of the bulk transport and thermodynamic properties of Pr124 single crystals.  From the combined torque and susceptibility data, the spins on the Pr sub-lattice were determined to be aligned predominantly along the $a$-axis. The ordering of the Pr ions at $T_{\rm N}$(Pr) also induces a peak in the heat capacity and a kink in the temperature dependence of the interchain resistivity $\rho_c(T)$. \cite{McBrien02} By studying the evolution of both the peak and the kink as a function of magnetic field and temperature, the magnetic phase diagram of the Pr ordering has been mapped out.

A strongly hysteretic discontinuity, characteristic of a spin-flop transition, was also revealed in both the heat capacity $C$ and the magnetic torque $\tau$ with a magnetic field of 10 Tesla aligned along the easy axis ({\bf H}$\|a$). At precisely the same field value, a kink was observed in the field derivative of the $c$-axis resistivity d$\rho_c$/d$H$, suggesting that the two phenomena are correlated. Significantly, the kink in  d$\rho_c$/d$H$ is non-hysteretic and persists to temperatures well above $T_{\rm N}$(Pr). Moreover, a similar feature has also been observed in non-magnetic Y124 \cite{Hussey98} and attributed to a field-induced dimensional crossover (FIDC) in the conduction electrons located on the CuO chains. \cite{Hussey02, Hussey98} The striking coincidence of the two field scales implies that the spin-flop transition in Pr124 is in fact {\it induced} by the FIDC of the chain carriers, presumably mediated via dominant and highly anisotropic RKKY interactions.

\section{Experimental}
\label{exp}

The Pr124 crystals were grown by a self-flux method in MgO crucibles under high-pressure oxygen gas of 11atm. \cite{Horii00} The susceptibility measurements were performed using a Quantum Design SQUID magnetometer. The $c$-axis resistivity measurements were carried out using a standard four-probe ac lock-in technique (at an excitation frequency of 77 Hz) in a (14 Tesla) superconducting magnet (in Bristol) and in a large (32 Tesla) dc magnet at the National High Magnetic Field Laboratory in Tallahassee. Torque magnetometry measurements were performed with a piezo-resistive cantilever, while heat capacity was measured as a function of temperature and field by an a.c.\ technique \cite{Carrington97} in fields up to 14~T. The torque and heat capacity measurements were carried out on sample $\sharp$1 (dimensions $0.22 \times 0.28 \times 0.022$ mm$^3$ and mass $\sim$ 9 $\mu$g) while the $c$-axis resistivity measurements were carried out on both sample $\sharp$1 and a second crystal, sample $\sharp$2 (dimensions $0.4 \times 0.15 \times 0.04$ mm$^3$). The susceptiblity data were taken on a third, much larger crystal, sample $\sharp$ 3, with a total volume of 1.8 mm$^3$.

For the $C(T, H)$ measurements, sample $\sharp$1 was attached to a flattened 12 $\mu$m diameter chromel-constantan thermocouple with GE varnish.  It was heated either with a modulated light source, or a very small resistive heater, at a frequency of (typically) 8~Hz. At this frequency $\omega T_{ac}$ was approximately independent of frequency $\omega$, indicating quasi-adiabatic conditions. \cite{Sullivan68} The field dependence of the thermocouple sensitivity was determined in a separate run by measuring the field dependent heat capacity of a similarly sized polycrystalline Cu sample.  The temperature of the main stage was measured with a Cernox thermometer.  The small size of the sample means that the contributions to the measured heat capacity from the thermocouple and glue are not negligible and are estimated to be $\sim$50\% of the total at $T=20$~K.  This also makes estimates of the absolute specific heat difficult and so the data are presented in arbitrary units. Absolute data for a large polycrystalline sample have been reported in Ref.~[10].

\section{Results}
\label{results}

\subsection{Magnetic Susceptibility}
\label{chi}

\begin{figure}
\includegraphics[width=8.5cm,keepaspectratio=true]{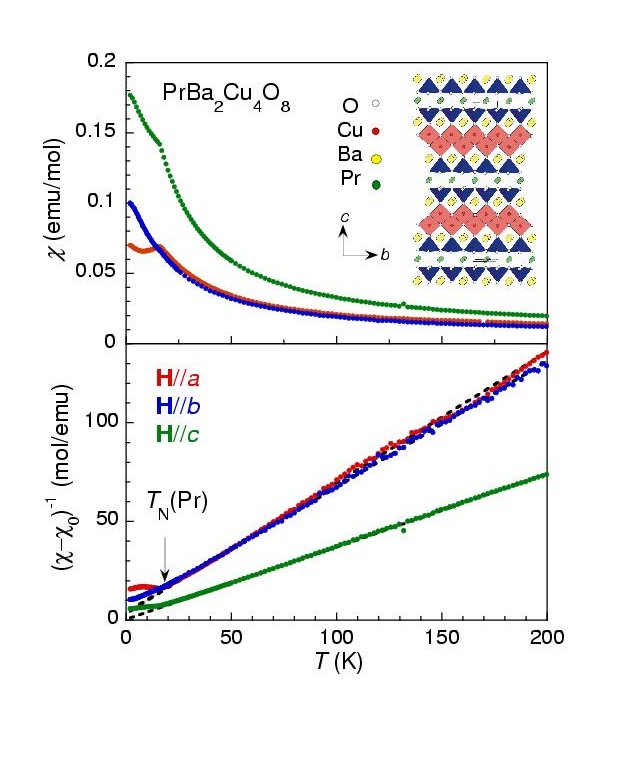}
\caption{(Color online) Top panel: Magnetic susceptibility of Pr124 (sample $\sharp3$) as a function of temperature for the magnetic field applied along the three crystallographic axes. Inset: Crystal structure of Pr124. The diamonds represent the CuO double chain network whilst the pyramids represent the 5 oxygen atoms surrounding each planar Cu atom. Bottom panel: Curie-Weiss law fits (dashed lines) to the same magnetic susceptibility data. The arrow indicates $T_{\rm N}$(Pr), the temperature at which the Pr ions order.} \label{Fig1}
\end{figure}

The top panel in Figure \ref{Fig1} shows the magnetic susceptibility along all three crystallographic axes between 2 K and 200 K in a magnetic field of 5 T. The main features of the data, in particular the strong Curie-Weiss Law dependence and the kink at 17 K, are similar to those reported previously in polycrystalline samples of Pr124 \cite{Yang97, Li99} and Pr123. \cite{Li89} The magnitude of the susceptibility for the field applied along the $c$-axis is roughly twice that for in-plane fields. Such anisotropy is commonly observed in layered cuprates with divergent susceptibilities and is attributed to anisotropic $g$ factors. \cite{Terasaki92}

To further analyze the susceptibility, the data are plotted in the bottom panel of Figure \ref{Fig1} as ($\chi - \chi_0)^{-1}$ versus temperature to compare with the Curie-Weiss law, $\chi(T) = \chi_0 + C/(T - \theta)$ where $C$ is a measure of the effective magnetic moment and $\theta$ is a parameter related to the interaction between the ions. Values obtained for the temperature independent $\chi_0$ values are 13.6, 8.76 and 12.3 x 10$^{-3}$ emu/mol for {\bf H}$\parallel$$a$, $b$ and $c$ respectively. The four main contributions to $\chi_0$ are: $\chi_P$ the Pauli paramagnetic term, $\chi_{\rm dia}$ the Landau diamagnetization, $\chi_{\rm core}$ the diamagnetic contribution of the core levels, and  $\chi_{\rm orb}$ the orbital Van Vleck paramagnetism. Of these, the first two, due to conduction electron effects and related to the density of states at the Fermi level, are expected to be the dominant terms. The other terms are typically of order 10$^{-5}$ or 10$^{-4}$ emu/mol, \cite{vanVleck} significantly less than the observed $\chi_0$. From the fits shown to the data in Figure \ref{Fig1} we can extract $\mu_{\rm eff}$, the effective moment per Pr ion, using the Curie-Weiss Law: $\chi(T) = n \mu_{\rm eff}^2/3k_B(T - \theta)$, where $n$ is the number of magnetic ions per mole. If we neglect the small Cu saturation moment, this yields $\mu_{\rm eff}$ values of 3.50$\mu_B$, 3.56$\mu_B$ and 4.70$\mu_B$ for each field orientation, broadly consistent with the value of 3.11 reported previously \cite{Yang97} and also with the magnetic moment of a free Pr$^{3+}$ ion ($\mu_{\rm eff}$ = 3.58$\mu_B$).

Above the ordering temperature, the spins on the Pr ions are thus behaving as independent or very weakly interacting moments. The ordering transitions are clearly seen in the data as deviations from the Curie-Weiss Law behavior. Due to the directions along which the spins order at the transition, the effects of the ordering transition on the magnetic susceptibility are dependent on the orientation of the external field with respect to the crystallographic axes. The susceptibility, $\chi(T)$, shows a clear maximum at $T_{\rm N}$(Pr) when field is applied along the $a$-axis and a kink for field applied along the $c$-axis. For field applied along the $b$-axis, on the other hand, only a weak inflexion point is observed. From symmetry arguments, \cite{Foner59} this implies that the spins are aligned predominantly along the $a$-axis. As we shall show below, this identification is corroborated by angular magnetic torque measurements.

\subsection{Magnetic Phase Diagram}
\label{phase}

\begin{figure}
\includegraphics[width=8.5cm,keepaspectratio=true]{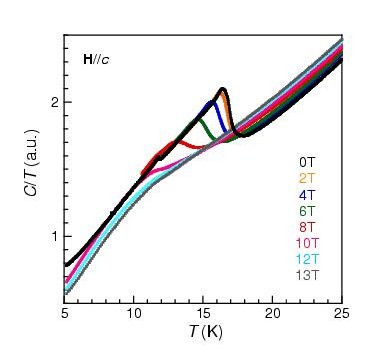}
\caption{(Color online) Temperature dependence of the heat capacity of sample $\sharp 1$ for different field strengths with {\bf H}$\parallel$$c$. In this sample, a second smaller peak resulting from some unknown minority phase is detected at 12 K in zero field.  Note that this peak is rapidly suppressed with applied field and vanishes for fields $>$ 6 T.} \label{Fig2}
\end{figure}

Figure \ref{Fig2} shows the heat capacity of a Pr124 single crystal ($\sharp$1) between 5 K and 25 K for {\bf H}$\parallel$$c$. In zero-field, there is a non-hysteretic peak in $C/T$ at $T$ = $T_{\rm N}$(Pr) = $17\pm0.1$ K (taking the mid-point of the initial rise), indicative of a second-order phase transition. As the strength of the external field is increased, the main anomaly in $C/T$ shifts to lower temperature, decreases in size and broadens considerably. For $\mu_0H\gtrsim 12$ T, the position of the transition is difficult to discern. Similar data (not shown) were obtained in the orthogonal field orientations, {\bf H}$\parallel$$a$ and {\bf H}$\parallel$$b$. Taking the mid-point of the rise as the magnetic ordering temperature of the Pr ions at each field, the magnetic phase diagram of Pr124 can thus be constructed. Similar measurements were performed for the other two field orientations, {\bf H}$\parallel$$a$ and {\bf H}$\parallel$$b$ (not shown) to complete the phase diagram, which is shown in Figure \ref{Fig6} and discussed in more detail below.

\begin{figure}
\includegraphics[width=8.5cm,keepaspectratio=true]{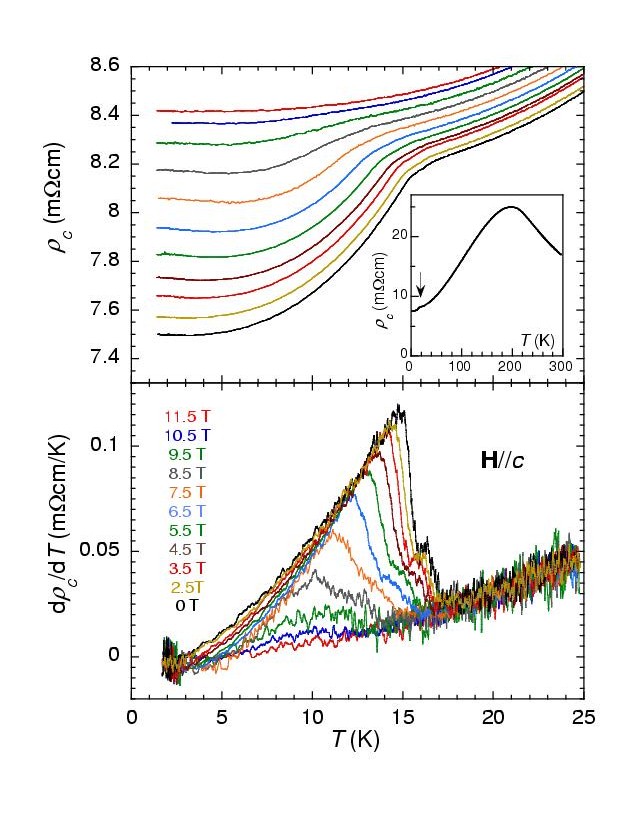}
\caption{(Color online) Top panel: Temperature sweeps of the $c$-axis resistivity of sample $\sharp$2 at different magnetic field strengths for {\bf H}$\parallel$$c$. Inset: Zero-field $\rho_c(T)$ curve for the same single crystal. The arrow indicates the kink due to the antiferromagnetic ordering of the Pr ions. Bottom panel: d$\rho_c$/d$T$ for the same curves shown above. The mid-point of the initial rise is taken to be $T_{\rm N}$(Pr) for each field strength.} \label{Fig3}
\end{figure}

As shown in the inset to the top panel in Figure \ref{Fig3}, the zero-field magnetic ordering of the Pr ions at $T_{\rm N}$(Pr) also manifests itself as a small kink in $\rho_c(T)$. \cite{McBrien02} The top panel of Figure \ref{Fig3} shows $\rho_c(T)$  for sample $\sharp$2 as a function of differing field strength directed along the $c$-axis. As the field is increased, the kink shifts to lower temperatures. At sufficiently high fields ($\mu_0H >$ 11.5 T), the kink is no longer visible, suggesting that the ordering has become completely suppressed. No corresponding kink is observed for {\bf I}$\parallel$$a$ or {\bf I}$\parallel$$b$ (except in more disordered crystals that exhibit resistivity upturns at low $T$, \cite{Narduzzo07, Rad07}) suggesting that the dominant spin-charge coupling is along the $c$-axis.

As is often found for itinerant magnetic systems, \cite{Fisher68} the evolution of the $C/T$ anomaly with field is mirrored in the temperature derivative d$\rho_c$/d$T$. Plotting the derivatives of the resistivity curves, d$\rho_c$/d$T$, as shown in the bottom panel of Figure \ref{Fig3}, we can see this suppression effect more clearly. The ordering of the Cu moments in the CuO$_2$ planes at 220 K\cite{Li99} also manifests itself as a minimum in the derivative $d\rho_c/dT$ (not visible in the inset of Fig.\ref{Fig3} - for more details, please refer to Fig.~\ref{Fig3} of Ref.~[6]).

\begin{figure}
\includegraphics[width=8.5cm, keepaspectratio=true]{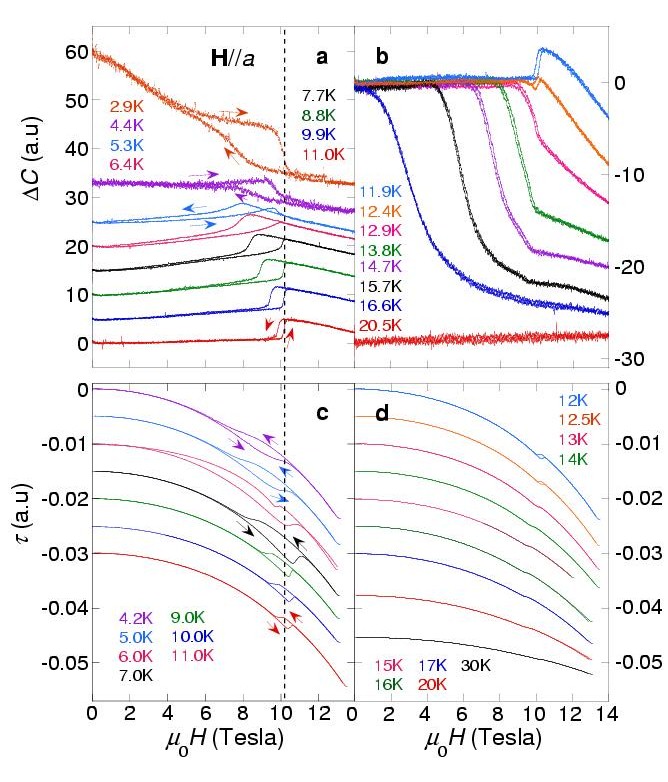}
\caption{(Color online) The magnetic field dependence ({\bf H}$\parallel$$a$) of the a),b) heat capacity and c),d) torque magnetization of sample $\sharp$1 at various temperatures, as marked, above (right panels) and below (left panels) $T$ = 11 K. The direction of the variation of the field is indicated by arrows in certain cases. The vertical dashed line is included to highlight the $T$-independence of the step in $C/T$ and $\tau$ on the increasing arm of the field sweep.} \label{Fig4}
\end{figure}

A striking new feature of the phase diagram is revealed upon performing field sweeps of the specific heat with {\bf H}$\parallel$$a$. Figure \ref{Fig4} (panels a) and b)) shows $C(H)$ sweeps carried out on sample $\sharp$1 at fixed temperatures between 3 K and 20 K, the temperature range of interest. Below $T$ = 12.5 K, $C(H)$ develops a step-like feature around $\mu_0H$ $\sim$ 10.2 T. The width of the hysteresis grows rapidly as the temperature is reduced further, though the onset of hysteretic behavior on increasing the field is relatively insensitive to temperature, as indicated by the vertical dashed line in Fig.~\ref{Fig4}. For 12 K $< T < T_{\rm N}$(Pr), the hysteresis vanishes and $C(H)$ exhibits a smooth decline with increasing field. Above $T_{\rm N}$(Pr), $C(H)$ is flat and featureless. \cite{OtherPeak} Field sweeps in the other field directions ({\bf H}$\parallel$$b$ and {\bf H}$\parallel$$c$) did not show any hysteretic jumps.

In order to investigate further the nature of this transition, complementary torque magnetometry measurements were performed on the same sample with {\bf H}$\parallel$$a$, as shown in panels c) and d) of Figure \ref{Fig4}. Just as in the $C(H)$ sweeps, $\tau(H)$ exhibits a step-like feature at low $T$ around $\mu_0H$ $\sim$ 10.2 T that becomes increasingly hysteretic with decreasing temperature. For 12 K $< T < T_{\rm N}$(Pr), the jump in $\tau(H)$ is transformed into an inflexion point while above $T_{\rm N}$(Pr), $\tau(H)$ varies as $H^2$, corresponding to a linearly-increasing magnetization. Although a small kink is still visible in $\tau(H)$ above $T_{\rm N}$(Pr), it slowly diminishes with increasing temperature. The similarity in the $C(H)$ and $\tau(H)$ sweeps, both in terms of the width and location of the hysteresis is clear and confirms the magnetic nature of the first-order transition. Consistent with other uniaxial antiferromagnets, \cite{Neel32, Rohrer75} we attribute this hysteretic feature to a spin-flop transition to a canted antiferromagnet state. We reiterate that the location of the spin-flop, as defined by the position of the step on the increasing field sweep, is almost independent of temperature.

\begin{figure}
\includegraphics[width=8.5cm,keepaspectratio=true]{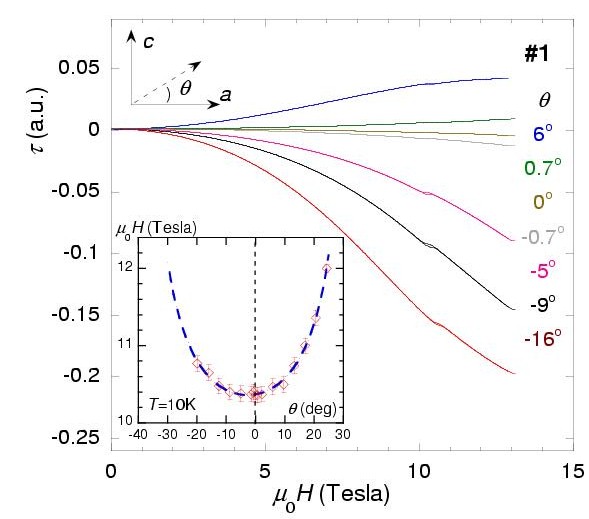}
\caption{(Color online) Torque magnetization of Pr124 (sample $\sharp$1) as a function of magnetic field at different orientations within the  $ac$-plane. The $\theta$ values represent the angle away from the easy axis {\bf H}$\parallel$$a$. Inset: Angle-dependence of the onset field for the step in $\tau(H)$ (for increasing field strengths).}
\label{Fig5}
\end{figure}

Figure \ref{Fig5} shows the variation of the field dependence of the torque magnetization for different angles within the $ac$-plane. The torque signal is found to be minimize when the magnetic field is oriented along the $a$-axis. Moreover, as shown in the inset to Fig.~\ref{Fig5}, the onset field for the step in $\tau(H)$ (for increasing field strengths) is also a minimum for {\bf H}$\parallel$$a$. Both these features confirm that the $a$-axis is the easy axis for the ordered Pr spins. This alignment of the spins is qualitatively consistent with that determined from the behavior of the magnetic susceptibility below $T_{\rm N}$(Pr) (see Figure \ref{Fig1}) but conflicts with earlier neutron diffraction measurements of Li {\it et al.} \cite{Li99} that suggested that the Pr ions order collinearly along the $c$-axis and contrasts with the situation in Pr123 where the Pr spins are found to be canted $\sim 35^{\rm o}$ out of the plane. \cite{Longmore96, Boothroyd97}

\begin{figure}
\includegraphics[width=8.5cm,keepaspectratio=true]{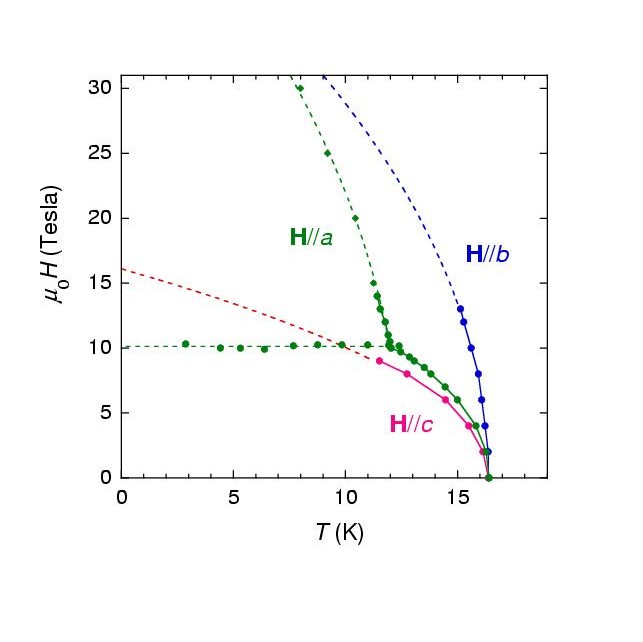}
\caption{(Color online) The magnetic phase diagram of the Pr ordering in Pr124, as determined from $c$-axis transport (solid diamonds) and heat capacity (solid circles) measurements. The dashed lines are fits to a mean-field like model. The horizontal dashed line is a guide to the eye.}
\label{Fig6}
\end{figure}

Figure \ref{Fig6} summarizes the resultant magnetic phase diagram of Pr124 as derived from the high-field $c$-axis transport measurements (shown as diamonds for {\bf H}$\parallel$$a$ and $|H| >$ 15 T) and heat capacity measurements (circles). The application of a magnetic field along any of the crystallographic axes suppresses the ordering anomaly though the suppression is most pronounced for {\bf H}$\parallel$$c$. The points corresponding to the spin-flop transition are defined for each temperature as the onset field during the up sweep of the hysteresis in the heat capacity (or torque) data (Figure \ref{Fig4}). The spin-flop transition is seen to occur at approximately 10 T for all $T <$ 12 K. As we shall discuss in the following section, this is precisely the same field ({\bf H}$\parallel$$a$) at which the conduction electrons in Pr124 undergo a field-induced dimensional crossover from 3D to 2D, suggesting an intimate relation between the two phenomena.

\subsection{Magnetic-field-induced dimensional crossover}
\label{dimension}

It was shown previously  that in high-quality Pr124 crystals, electrical resistivity at low $T$ is metallic in all three orthogonal directions and varies approximately as $T^2$, consistent with the development of a three-dimensional (3D) Fermi-liquid ground state. The resistive anisotropy at low $T$ is extremely large however ($\rho_a$:$\rho_b$:$\rho_c \sim$ 1000:1:3000), \cite{McBrien02} with a similar anisotropy in the ratio of the squares of the hopping energies. At high $T$, the interchain resistivities $\rho_a(T)$ and $\rho_c(T)$ have maxima around $T$ = 150 K above which d$\rho_{a,c}$/d$T < 0$. \cite{McBrien02} These maxima can be interpreted either as a thermally-induced 3D to 1D dimensional crossover in the electronic ground state \cite{Giamarchi} or the emergence of a second conduction channel (e.g. the insulating CuO$_2$ planes) at higher temperatures. \cite{Hussey98b, Gutman07}

The Fermi surface of Pr124 is believed to consist of two pairs of corrugated sheets extending normal to the reciprocal space axis ${\mathbf k}_{b}$. Within a simple tight-binding picture, the $c$-axis dispersion (for a single chain) is $\mathcal{E}=-2t_{\perp}^{c}$cos$(k_{c}c)$. For {\bf H}$\parallel$$a$, the dominant Lorentz force $e\mu_0[{\bf v}_{F}\times{\bf H}]=e\mu_0v_{F}H{\bf \hat{c}}=\hbar d{\bf k}_{c}/dt$ causes carriers to traverse the Fermi sheet along $k_{c}$. The sinusoidal corrugation then gives rise to an oscillatory component to the $c$-axis velocity $v_{\perp}^{c}$ and hence to a real-space sinusoidal trajectory with amplitude $A_{c}=2t_{\perp}^{c}/e\mu_0v_{F}H$. Thus $A_{c}$ shrinks as $H$ increases until eventually at $\mu_0H_{\rm cr}^{a}=2t_{\perp}^{c}/ev_{F}c$, $A_{c}=c$ and the charge carriers become confined to a single plane of coupled chains. Note that $\mu_0H_{\rm cr}$ is independent of the scattering rate $1/\tau_0$. This magnetic field-induced dimensional crossover is a classic signature of a quasi-1D conductor. \cite{Hussey98, GorkovLebed, Behnia95}

\begin{figure}
\includegraphics[width=7.5cm, keepaspectratio=true]{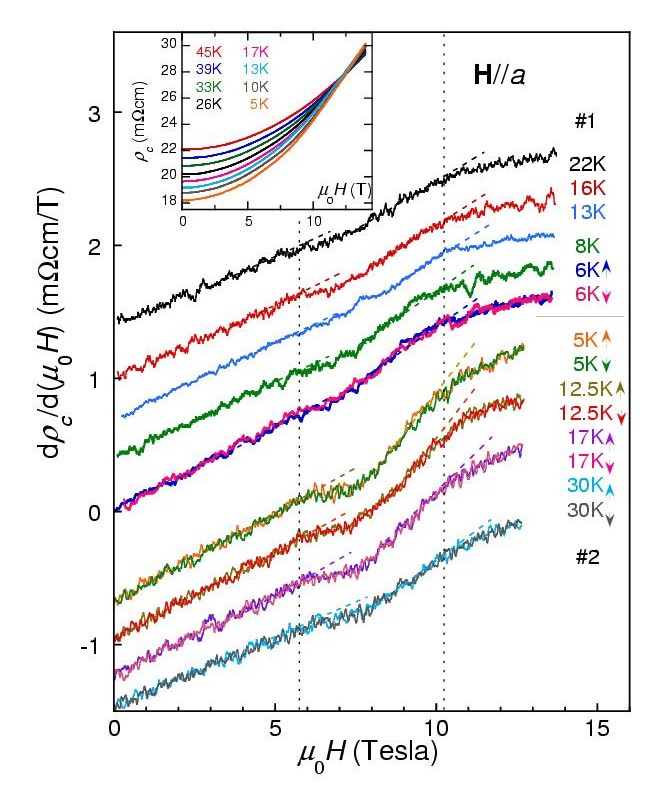}
\caption{(Color online) d$\rho_c$/d$H$ versus $H$ for samples $\sharp$1 and $\sharp$2 at different temperatures with {\bf H}$\parallel$$a$ (offset for clarity). The vertical dashed lines locate the kinks in d$\rho_c$/d$H$ that signify the (dimensional) crossover field for the individual chain sheets. The other dashed lines are guides to the eye. Inset: $\rho_c$ versus $H$ for sample $\sharp$1 showing the metallic to non-metallic crossover in $\rho_c(T)$ at $\mu_0H$ $\sim$ 12T ({\bf H}$\parallel$$a$).} \label{Fig7}
\end{figure}

The main panel in Fig.~\ref{Fig7} shows d$\rho_c$/d$H$ versus $H$ at various temperatures above and below $T_{\rm N}$(Pr) for both crystals $\sharp$1 and $\sharp$2 with {\bf H}$\parallel$$a$. There are two features in the field derivative of both crystals (marked by vertical dashed lines), one at 6 Tesla, the other at $\sim$ 10 Tesla. Both features are characterized by a drop in d$\rho_c$/d$H$ (the changes being more pronounced in $\sharp$2) that occurs at roughly the same field value for all $T$. Other Pr124 crystals with differing impurity levels have been found to exhibit similar features at precisely the same field strengths (see, e.g. Figure 4 of Ref.~[7]). As discussed above, such insensitivity of $\mu_0H_{\rm cr}$ to the strength of both the elastic and inelastic scattering rates is a notable feature of the Gorkov-Lebed theory of the FIDC. \cite{Waku00}

The signatures of the FIDC in Y124 and Pr124 are two-fold: a reduced d$\rho_c$/d$H$ at $H_{\rm cr}^{a}$ and a change in the $T$-coefficient of $\rho_c(T)$, $\alpha = (1/\rho)$(d$\rho$/d$T$), from positive to negative (see inset to Fig.~\ref{Fig7}). \cite{Hussey98, Hussey02} The two signatures do not necessarily occur at the same field value however. Indeed, while the kinks in d$\rho_c$/d$H$ are found to be identical in {\it all} crystals measured to date, the metal to non-metal crossover in $\rho_c(T)$ has been observed to occur at fields less than 10 Tesla, \cite{Hussey98} at fields equal to 10 T \cite{Narduzzo06} and, as seen here, at fields above $H_{\rm cr}^{a}$. In order to understand why this is so, it is important to note that while the residual $c$-axis resistivity is proportional to 1/$\tau_0$, the interchain magnetoresistance is proportional to $\tau_0^2$. Thus, in contrast to the kink field(s), the crossover field from metallic to non-metallic $\rho_c(T)$ depends on 1/$\tau_0$. Nevertheless, both phenomena are related to the same FIDC.

Due to hybridization effects, the separate sheets of the double CuO chain unit in Pr124 are expected to have slightly different $k_F$ values and likewise different $c$-axis warpings, as seen in the calculated electronic band structure of isostructural Y124. \cite{Yu91} With regards to the theory, we therefore attribute the lower kink field to the sheet with the smallest $t_{\bot}$ whilst the higher kink field corresponds to $H_{\rm cr}^{a}$ for the dominant, more highly warped chain sheet. The $t_{\bot}$ values thus obtained are consistent with the temperature scales at which each feature becomes smeared out. \cite{Hussey02} Above $H_{\rm cr}^{a}$ for the second sheet, the chain carriers will be confined to the $ab$-plane and the $T$-dependence of $\rho_c(T)$ becomes non-metallic. Note that a similar FIDC also occurs in the reciprocal configuration, i.e. for {\bf I}$\parallel$$a$ and {\bf H}$\parallel$$c$, though at a much higher field $\mu_0H_{\rm cr}^{c}$ = 60 T since there $\mu_0H_{\rm cr}^{c}=2t_{\perp}^{a}/ev_{F}a$ and $a \sim c/3$. \cite{Narduzzo06}

\section{Discussion}
\label{discussion}

\begin{figure}
\includegraphics[width=7.5cm, keepaspectratio=true]{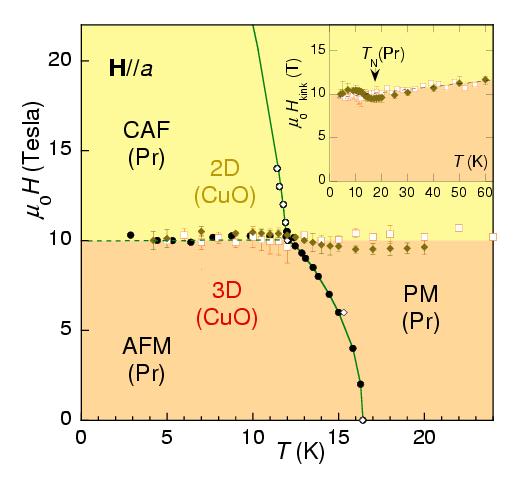}
\caption{(Color online) Magnetic phase diagram of the Pr ordering in Pr124 for {\bf H}$\parallel$$a$ from $C(T, H)$ (solid circles), plotted together with the positions of the kinks seen in d$\rho_c$/d$H$ (open symbols) and $\tau(T, H)$ (solid diamonds). The labels AFM, PM and CAF stand for the antiferromagnetic phase, the paramagnetic Curie-Weiss phase, and the canted AFM phase respectively. The different shaded regions represent the 3D-2D FIDC in the chain conduction electron sub-system. Inset: $H_{\rm kink}(T)$ from $\rho_c(T, H)$ (open squares) and $\tau(T, H)$ (solid diamonds) over an extended temperature range. The arrow indicates the position of $T_{\rm N}$(Pr) in zero-field.} \label{Fig8}
\end{figure}

Figure \ref{Fig8} summarizes the resultant magnetic phase diagram of Pr124 for {\bf H}$\| a$ as derived from $\Delta C(H, T)$ (solid circles), $\tau(H, T)$ (solid diamonds) and d$\rho_c$/d$H(T)$ (open squares). The points corresponding to the spin-flop transition are defined as the onset field during the up sweep of the hysteresis in either $\Delta C(H)$ or $\tau(H)$ data (Fig. \ref{Fig4}). The spin-flop transition is seen to occur at approximately 10 T for all $T <$ 12 K. Consistent with other uniaxial antiferromagnets, the spin-flop transition ends at a triple point (at $T \sim$ 12 K), below which there is a significant enhancement of the AFM ordering of the Pr ions.

As stressed throughout, the second kink feature in d$\rho_c$/d$H$ at $\mu_0H_{\rm cr}$ = 10.2 Tesla occurs at precisely the same field strength as the step in $C/T(H)$ and $\tau(H)$. This striking coincidence between the FIDC and the hysteretic thermodynamic transition is the main result of this article. Given how closely d$\rho_c$/d$T$ mirrors the Pr ordering anomaly in $C/T$ for {\bf H}$\parallel$$c$ (Fig.~\ref{Fig3}), the lack of hysteresis in d$\rho_c$/d$H$ strongly suggests that whilst the two {\bf H}$\parallel$$a$ features are coupled, they correspond to different physical phenomena. Were the kink in d$\rho_c$/d$H$ that occurs at 10 Tesla on the up-sweep induced by the canting of the Pr spins, then either a similar anomaly would be observed in d$\rho_c$/d$H$ on the down-sweep at exactly the same point where the AFM alignment is recovered, or there would be no response at all. In this case however, there is a kink at 10 Tesla on {\it both} the up- and the down-sweeps. This lack of hysteresis, coupled with the persistence of the kink in d$\rho_c$/d$H$ to temperatures way beyond $T_{\rm N}$(Pr) and the observation of similar behavior in non-magnetic Y124 \cite{Hussey98} constitutes strong evidence therefore that it is the dimensional crossover in the charge sector (at 10 Tesla) that drives the spin-flop transition in the Pr sub-lattice, rather than the other way around.

Within this scenario, the survival of the kink in $\tau(H)$ to temperatures above $T_{\rm N}$(Pr), where it continues to track the kink in d$\rho_c$/d$H(T)$ (see inset to Fig.~\ref{Fig8}), can be attributed to a change in the anisotropy of the Pauli susceptibility induced by the FIDC. The only viable alternative scenario would be to attribute the resistive kink to some residual spin-flop dynamics occurring on the planar Cu moments (which order antiferromagnetically at $T \sim$ 200 K) \cite{Li99}. However, we note that there is no experimental signature of a first-order jump in either $C(H)$ or $\tau(H)$ beyond $T$ = 12 K (see Fig. \ref{Fig4}) and to the best of our knowledge, no such spin-flop transition has ever been detected in the (magnetically ordered) Y-based cuprates.

This symbiosis between the dimensionality of the conduction electrons and the re-arrangement of the Pr spins suggests the presence of a strong, RKKY-type coupling between the two sub-systems. The ordering temperature of the localized Pr moments is determined largely by the $c$-axis exchange interaction, both direct (super)exchange $J_S^c$ between neighboring ions and indirect exchange $J_{\rm RKKY}^c$ between the Pr ions and the conduction electrons. Since the Pr cations are far apart ($c$ = 13.6$\AA$), the superexchange term is expected to be rather small and thus may not play the dominant role here. The RKKY interaction, on the other hand, decays slowly with distance. It is proportional to the hybridization between the localized and itinerant carriers (or between the single-ion wave-function and that of the itinerant carriers). Hence, the magnitude of the individual matrix elements of $J_{\rm RKKY}$ will depend on the individual hopping (kinetic) energy terms of the conduction electrons, via their Pauli paramagnetic susceptibility, and hence ultimately, on the topology of their Fermi surface. \cite{Zhou10}

At the FIDC, $t_{\bot}^c$ of the chain carriers is effectively renormalized to zero. This corresponds to an effective change in the spatial distribution of the wave-function of the itinerant carriers. Na\"{\i}vely one would expect this to affect the degree of hybridization between the two sub-systems (localized moments and itinerant carriers) and thus to modulate the strength of the RKKY interaction. This renormalization of the $c$-axis component of the exchange interaction matrix $J_{\rm RKKY}^c$ (towards zero) at $H_{\rm cr}^{a}$ then destabilizes the spin configuration and induces the spin-flop transition. While the Pr moments may well have undergone a similar spin-flop transition without the intervention of the conduction electrons, it would appear that the precise field at which this occurs is set predominantly by the dimensional crossover.

\section{Conclusions}
\label{conclusions}

In summary, through a combination of electrical resistivity, magnetoresistance, specific heat and magnetization measurements, performed for the first time on single crystalline samples, we have been able to generate a coherent picture of both the geometry and the dynamics of the antiferromagnetic ordering of the localized Pr spins in Pr124. The Pr spins are found to be aligned predominantly along the $a$-axis, in contrast to previous claims from neutron diffraction measurements of  collinear ordering of the Pr spins along the $c$-axis. \cite{Li99}

In addition, a spin-flop transition has been found whose location in the magnetic phase diagram is coincident with a field-induced charge confinement of the conduction electrons, implying an intimate relation between the two phenomena. We speculate that the spin-flop transition and the corresponding enhancement of the AFM ordering of the Pr ions for {\bf H}$\parallel$$a$ at 10 T are a direct manifestation of the anisotropic RKKY interaction in Pr124, with the ($c$-axis) interaction strength $J_{\rm RKKY}$ being substantially renormalized at the crossover field $H_{\rm cr}^{a}$. To the best of our knowledge, this finding represents the first report of a spin-flop transition being driven by a dimensional crossover in the electronic state of the charge carriers. It also implies that the field-induced real space confinement of the chain carriers in Pr124 is a genuine physical effect with instantaneous consequences. Further investigation is envisaged to understand fully the dynamics of the relation between the spin-flop transition and the FIDC. In this regard, it would certainly be interesting and informative to study electron spin resonance of the Pr spins to see if the Zeeman splitting of the magnetic sub-levels displays a non-linear response at or near $H_{\rm cr}^{a}$.

\begin{acknowledgments}
The authors would like to acknowledge technical assistance from L. Balicas, B. Fauqu\'{e}, M. N. McBrien and O. J. Taylor and helpful discussions with A. Boothroyd, R. Coldea and N. Shannon. This work was supported by the EPSRC (UK), the Royal Society and a cooperative agreement between the NSF and the state of Florida (USA).
\end{acknowledgments}


\begin{thebibliography}{32}
\bibitem{OtherPeak} Note that none of these features can be attributed to the small $C(T)$ anomaly seen in zero-field and attributed to some impurity phase since its intensity is much too weak and was found to disappear in fields abvove 6 T.
\end{thebibliography}


\end{document}